\documentclass[a4paper, twocolumn, 12pt]{article}

\usepackage[english]{babel}
\usepackage[utf8x]{inputenc}
\usepackage[T1]{fontenc}

\usepackage[font={small}]{caption}

\usepackage[a4paper,top=3cm,bottom=3cm,left=1.5cm,right=1.5cm,marginparwidth=1.75cm]{geometry}

\usepackage{amsmath}
\usepackage{graphicx}
\usepackage{float}
\usepackage{placeins}
\usepackage{fancyhdr}
\usepackage[colorlinks=True, citecolor = blue, urlcolor=cyan]{hyperref}

\usepackage[format=plain,margin=0pt,font={small},labelfont={bf},
labelsep=endash]{caption}
\DeclareCaptionLabelSeparator{bar}{ | }
\usepackage{layouts}

\title{\vspace{-3.5cm}\textbf{Open(G)PIAS:} An open source solution for the construction of a high-precision Acoustic-Startle-Response (ASR) setup for tinnitus screening and threshold estimation in rodents}
\author{Richard Gerum$^{1*}$, Hinrich Rahlfs$^{2,3*}$, Matthias Streb$^{2,3*}$, Patrick Krauss$^{2}$,\\ Claus Metzner$^{1}$, Konstantin Tziridis$^{2}$, Michael Günther$^{3}$,\\ Holger Schulze$^{2}$, Walter Kellermann$^3$, Achim Schilling$^{2}$}
\date{}

\begin{document}
\onecolumn
\maketitle
~\newline
\vspace{-1cm}
\begin{flushright}
\begin{minipage}[t]{0.95\textwidth} 
\textit{$^1$ Department of Physics, Center for Medical Physics and Technology, Biophysics Group, Friedrich-Alexander University Erlangen-Nürnberg (FAU), Erlangen, Germany\\
~\newline
$^2$ Experimental Otolaryngology, ENT-Hospital, Head and Neck Surgery, Friedrich-Alexander University Erlangen-Nürnberg (FAU), Erlangen, Germany\\
~\newline
$^3$ Multimedia Communications and Signal Processing, Friedrich-Alexander University Erlangen-Nürnberg (FAU), Erlangen, Germany\\
~\newline
$^*$ authors contributed equally to this work} 
\end{minipage}
\end{flushright}

\vspace{1cm}
\begin{abstract}
\noindent The acoustic startle reflex (ASR) that can be induced by a loud sound stimulus can be used as a versatile tool to, e.g., estimate hearing thresholds or identify subjective tinnitus percepts in rodents. These techniques are based on the fact that the ASR amplitude can be suppressed by a pre-stimulus of lower, non-startling intensity, an effect named pre-pulse inhibition (PPI). For hearing threshold estimation, pure tone pre-stimuli of varying amplitudes are presented before an intense noise burst serving as startle stimulus. The amount of suppression of the ASR amplitude as a function of the pre-stimulus intensity can be used as a behavioral correlate to determine the hearing ability. For tinnitus assessment, the pure-tone pre-stimulus is replaced by a gap of silence in a narrowband noise background, a paradigm termed GPIAS (gap-pre-pulse inhibition of the acoustic startle response). The rationale of this approach is that the presence of a tinnitus percept leads to a masking of that gap (``filling in'' hypothesis) leading to a reduced PPI in the affected frequency range. 
A proper application of these paradigms depend on a reliable measurement of the ASR amplitudes, an exact stimulus presentation in terms of frequency and intensity. Here we introduce a novel open source solution for the construction of a low-cost ASR setup for the above mentioned purpose. The complete software for data acquisition and stimulus presentation is written in Python 3.6 and is provided as an anaconda package. Furthermore, we provide a construction plan for the sensory system based on low-cost hardware components. Exemplary data show that the ratios (1-PPI) of the pre and post trauma ASR amplitudes can be well described by a lognormal distribution being in  good accordance to previous studies with already established setups. Hence, the open access solution described here will help to further establish the ASR method in many laboratories and thus facilitate and standardize research in animal models of tinnitus or hearing loss.  
\end{abstract}

\twocolumn
\section*{Introduction}
A behavioral paradigm that can be used to assess hearing abilities in animal models (i.e. behavioral audiometry) without the necessity to apply time-consuming conditioning paradigms is the so called pre-pulse inhibition of the acoustic startle reflex (PPI of ASR or PIAS).
The startle reflex is induced by an intense stimulus such as a loud tone. This reflex, processed in the brainstem, is an evolutionary adaptation to prevent subjects from harm, e.g., by sudden predators \cite{Koch1999}. The reaction to a loud tone, the acoustic startle reflex (ASR), can be modulated by a variety of different factors such as drug treatment, pathological conditions, or presentation of pre-stimuli \cite{Koch1999}. A decrease of the ASR amplitude caused by any kind of pre-stimulus is called pre-pulse inhibition (PPI) and indicates that the stimulus has actually been perceived \cite{Fendt2001}. Hence, e.g. hearing thresholds can be determined by acoustic pre-stimuli such as pure-tones of varying intensity \cite{Tziridis, Walter2012}. Furthermore, in 2006, Turner and coworkers suggested a novel paradigm using the ASR as a tool for tinnitus screening \cite{Turner2006}. The paradigm is based on the fact that not only a tone or noise pre-stimulus can reduce the startle response but also a gap of silence embedded in band pass filtered noise can lead to a suppression of the ASR. This paradigm is called "Gap-Pre-Pulse Inhibition of the Acoustic Startle Reflex" (GPIAS). Thus, a potential tinnitus percept leads to a masking of this gap of silence and consequently results in a diminished decrease of the startle amplitudes (decrease of PPI) \cite{Turner2006}. It is advantageous that this paradigm does not depend on any pre-training of the animals such as classical conditioning (e.g. \cite{Jastreboff1988}). As it is less time consuming compared to conditioning paradigms, this paradigm is widely used in the tinnitus research community \cite{Kalappa2014,Shore2016,Krauss2016,Pienkowski2018}. Thus, the GPIAS paradigm and the evaluation procedures are continuously advanced and improved \cite{Longenecker2012,Schilling2017}.
However, setups for the recording of ASR responses for tinnitus screening as well as hearing ability estimation are still quite expensive. Our aim is to provide an open source solution for the construction of an ASR setup that is written in the interpreted programming language Python and is based on commercially available hardware components.

For clarification of the method, we also provide some testing results in our animal model, the Mongolian gerbil (\textit{Meriones unguiculatus}).

\section*{Materials and Methods}
\subsection*{Animals and Housing}
The Mongolian gerbils were housed in standard animal racks (Bio A.S. Vent Light, Ehret Labor- und Pharmatechnik, Emmendingen, Germany) in groups of 3--4 animals with free access to water and food at a room temperature of 20--25\,$^\circ$C under a 12h/12h dark/light circle. The care of the animals was approved by the state of Bavaria (Regierungspräsidium Mittelfranken, Ansbach, Germany, No. 54-2532.1-02/13). Exemplary measurements were recorded using animals aged ten to twelve weeks purchased from Janvier Laboratories Inc.

\subsection*{Software}
The complete software for the ASR setup is written in Python 3.6 \cite{Rossum1995}, an interpreted programming language optimized for scientific purposes. For maximum efficiency several open source libraries are used. 
For numerical operations, such as matrix operations, the Numpy library is used \cite{VanderWalt2011}. Further complex mathematical operations, such as signal filter functions, are implemented using the SciPy package \cite{Olifant2007}. The data are visualized using the Matplotlib library \cite{Hunter2007}.

\begin{figure}[htb]
\centering
\includegraphics{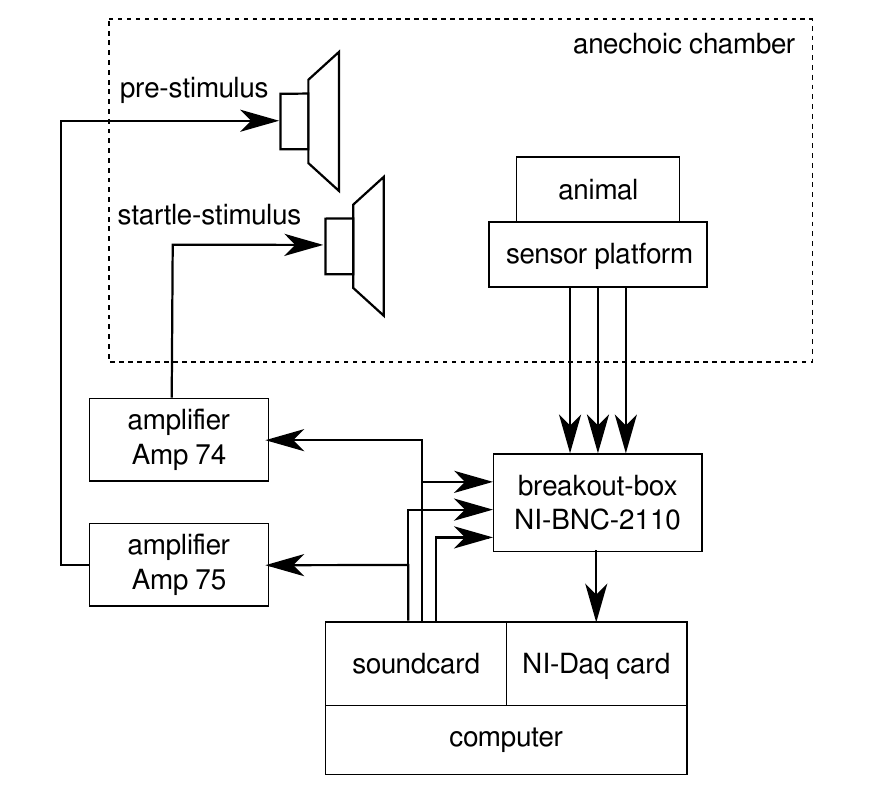}
\caption{\textbf{ASR measurement setup}\newline
Stimuli are presented via two different loudspeakers, one for the pre-stimulus and one for the startle stimulus, both placed in an anechoic chamber. Both loudspeakers are connected via an amplifier (Amp 74 and Amp 75) to the soundcard of the computer. The animal is restrained in an acrylic tube, located on a sensor platform in the anechoic chamber. ASR amplitudes are measured via an acceleration sensor. Sound and accelerations are recorded using a data acquisition card connected to the PC over a breakout box. Our open source ASR program (Python) controls stimulus application and measurement recording.}
\label{fig:Hardware_Config}
\end{figure}

\section*{Setup}

The setup is located in an anechoic chamber (Industrial  Acoustics  Company  GmbH, Niederkrüchten, Germany) and consists of two main parts: the stimulation hardware and the recording hardware (see Fig. \ref{fig:Hardware_Config}). The animal is restrained in an acrylic tube (different inner diameter: 27\,mm, 37\,mm, or 42\,mm, depending on the size of the animal) placed on a sensor platform with an integrated acceleration sensor (ADXL 335 on GY 61 board). The animal is acoustically stimulated using two different loudspeakers. A broadband two-way loudspeaker (Canton Plus XS.2) presents the pre-stimuli such as narrowband noise or pure tones of low amplitudes. To protect the broadband loudspeaker from damage by loud stimuli, a second loudspeaker (Neo-25s, Sinuslive) is used to present the startle stimuli. The startle stimuli causing the animal to twitch (ASR-amplitudes) are 20\,ms noise bursts, with an intensity of 115 dB SPL, proven to be the optimal stimulus to induce an ASR response (cf. \cite{Turner2006}). This same startle stimulus is used for all paradigms (e.g. threshold or GPIAS measurements). All stimuli are presented via a soundcard (Asus Xonar STX II) connected to two pre-amplifier (Amp 75 for pre-stimuli, Amp 74 for startle stimuli, Thomas Wulf, Frankfurt). 

The recording and digitization of the analog signal of the acceleration sensor (ADXL 335 on GY 61 board, Robotpark) is performed by a data acquisition card (PCIe-6320, National Instruments) connected to the sensor via a breakout box (NI-BNC 2110). The synchronization of startle stimulus onset and ASR response is assured by a trigger pulse generated by the soundcard, sent directly to the data acquisition card. The rising edge of the trigger-pulse is used to align the measured acceleration values to the startle stimulus onset.

\begin{figure}[htb]
\centering
\includegraphics{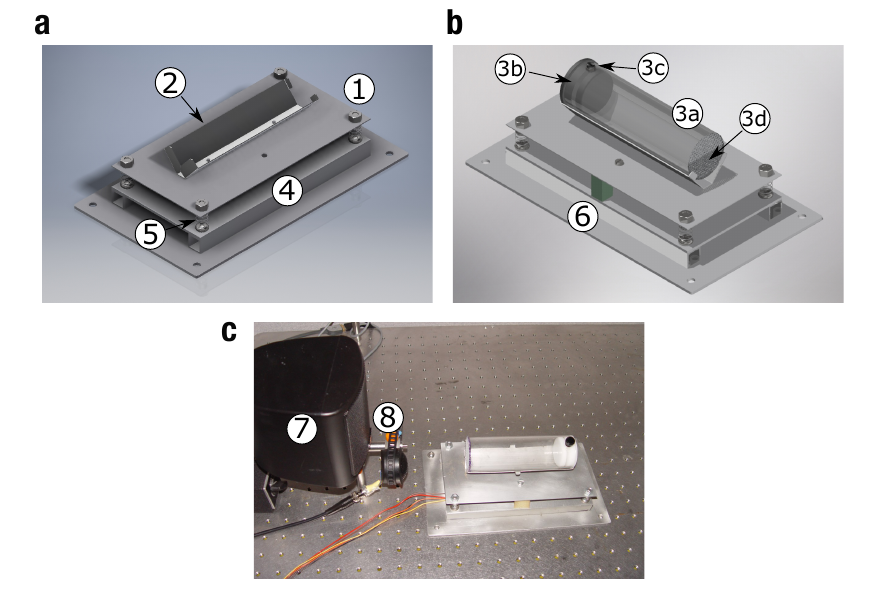}
\caption{\textbf{Sensor platform for ASR quantification}\newline \textbf{a}, Sensor platform (without animal restrainer). \textbf{b}, Sensor platform (with animal restrainer). \textbf{c}, Photo of the sensor platform in front to the loudspeakers. The upper plate (1) of the sensor platform holds the mount (2) for the animal restrainer (3).  The restrainer consists of an acrylic tube (3a) fixed in the mount, closed by a 3D printed plastic cap (3b), fixed with a screw (3c), at the rear end, and a wire-mesh (3d) at the front end, facing the loudspeaker. 
The acceleration sensor is fixed underneath the upper plate. This plate is flexibly mounted on the lower plate (4) via four springs (5). The lower plated is screwed to the vibration isolated table. Rubber foam (6) to prevents the sensor system from oscillating. The sensor platform is placed in front of the two speakers: pre-stimulus (7) and startle-stimulus loudspeaker (8).}
\label{fig:Sensors}
\end{figure}

%\FloatBarrierr

The sensor platform consists of two plates (cf. Fig. \ref{fig:Sensors}a1,4). The lower plate is fixed to the vibration isolated table (TMC, Peabody, MA, USA) by four screws. The upper plate is flexibly mounted on the lower plate by four springs, damped by two foam rubber blocks. The upper plate has a mount for the animal restrainer. This restrainer consists of an acrylic tube which is closed on both ends, with a wire mesh at the front end, facing the speakers, and a plastic cap (custom 3D print) at the rear end. The wire mesh causes no measurable distortions in the spectra of the acoustic stimuli and the cap prevents the animal from escaping the restrainer. The fixed mount ensures a constant distance between the animal and the loudspeakers.

As described above, ASR amplitudes are quantified by a three-way acceleration sensor, fixed underneath the upper plate of the sensor platform. When the animal twitches, as a response to the startle stimulus, the upper plate moves. This movement is quantified by the acceleration sensor. The ASR amplitude is
\begin{align}
A=\text{max}_t(a(t)\cdot\Theta(t)\cdot \Theta(150\,\text{ms}-t)),
\label{eq:forcevector}
\end{align}

where $\Theta$ is the Heaviside function, $t=0$ the begin of the startle stimulus and
\begin{align}
a(t)=\sqrt{(c_x\cdot a_{x(t)})^2+(c_y\cdot a_{y(t)})^2+(c_z\cdot a_{z(t)})^2},
\end{align}
with $a_{x(t)}, a_{y(t)},a_{z(t)}$, the measured acceleration in $x,y,z$ direction and $c_x, c_y, c_z$ the calibrated values, so that same force leads to same acceleration. 

\section*{Stimulation Software}
The software consists of several parts, a configuration module that allows to adapt the software to the given hardware setting, a protocol generator that allows to prepare the stimuli that should be presented, and a measurement module that applies the specified stimuli to the animal and records the responses.
\subsection*{Configuration}
The configuration module allows to specify details of the used hardware: The soundcard and sound driver as well as the channels for trigger pulse, pre-stimulus, and startle-stimulus can be specified.

The configuration module also allows to perform some calibration measurements, e.g., for synchronizing the output channels of the soundcard or equalizing the frequency response of the loudspeakers.

To calibrate possible latency shifts, a TTL pulse is presented in all selected output channels and the results are recorded using the data acquisition card. The program uses these measurements to determine the relative time shifts between the channels.

\begin{figure}[htb]
\centering
\includegraphics{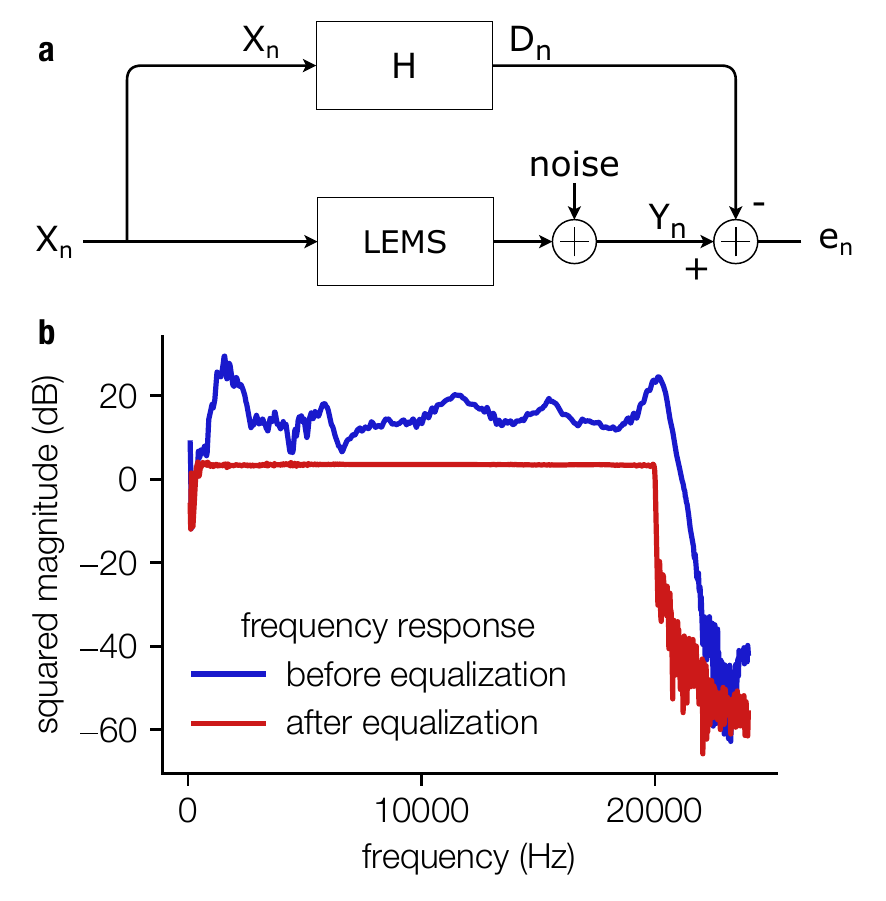}
\caption{\textbf{Equalizer Scheme}\newline
\textbf{a}, The block diagram of the system identification task; The output signal $X_n$ of the soundcard  is fed into the ``LEMS" (speaker) and recorded by the microphone (subject to additive measurement noise). The same signal is fed into the adaptive filter $H$ and the difference signal $e_n$ is used to adjust the filter coefficients. This iterative process leads to a filter that mimics the transfer function of the "unknown system". \textbf{b}, The frequency response of the overall system without equalization (blue) and with equalization (red). The frequency response is flattened in the range from 2\,kHz to 20\,kHz, where the microphone exhibits a flat frequency response.}
\label{fig:Equ}
\end{figure}

Since loudspeakers often exhibit a non-flat frequency response, it is desirable to correct these deviations using an equalizing filter.
To this end, a microphone with a flat frequency response is placed at the position of the animal in the measurement setup and the influence of the animal itself on the sound field is emulated by a piece of rubber foam.
The coefficients of the equalizer filter are determined by first identifying the loudspeaker-enclosure-microphone system (LEMS) and subsequently inverting its transfer function in the frequency domain.
The system identification task, depicted schematically in Fig.~\ref{fig:Equ}a, is solved by an adaptive linear filter using the NLMS algorithm \cite{Widrow1976}, which iteratively minimizes the power of the error signal between the adaptive filter output $D_n$ and the observed microphone signal $Y_n$ for a known excitation signal $X_n$.
To allow the identification of all frequencies of interest, the excitation signal should exhibit a constant power spectral density in the relevant frequency range as provided, e.g. by a white noise sequence or pseudo-random maximum-length sequences \cite{Rife1989}.
Finally, the time-domain coefficients of the equalizer filter are obtained by frequency bin-wise inversion of the identified transfer function over the desired frequency range and subsequent application of the inverse discrete Fourier transform (DFT), realized by a Fast Fourier Transform (FFT).
By convolving the loudspeaker driving signals with the equalization filter before playback, a flat frequency response for the overall system, i.e. the cascade of the equalization filter and the loudspeaker-enclosure-microphone system can be achieved as depicted in Fig.~\ref{fig:Equ}b.

\subsection*{Protocols}
The protocol module allows to define measurement protocols for different paradigms. Currently two types of protocols are implemented: hearing threshold measurements (based on PIAS) and GPIAS measurements, but the system is flexible to allow the future implementation of additional protocols.

All measurement sessions consist of multiple trials, each ending with a startle-stimulus. At the beginning of each measurement session, five trials without pre-stimulus are presented to prevent adaptation effects in the response amplitudes of the animal during the subsequent trials.

For hearing threshold determination, some trials have a pre-stimulus in the form of a short (40\,ms) pure tone 100\,ms before the startle stimulus. For the protocol, a frequency range (with octave, 1/2 octave, or 1/4 octave steps), a sound pressure level range, and a trial repetition count can be specified. The number of trials with pre-stimulus is the same as the number of trials without a pre-stimulus.

For the GPIAS paradigm, each trial has a noise presented before the startle-stimulus. The noise can be broadband or a frequency band around a center frequency. For each measurement different center frequencies can be specified, as well as the number of measurement repetitions. The noise can be interrupted by a short (50\,ms, flattened with 20\,ms sin$^2$-ramps) gap of silence. The number of trials with a gap is the same as the number of trials without a gap.

The trials in each protocol are randomized to prevent habituation effects of the animal.

\begin{figure}[htb]
\centering
\includegraphics[width=0.4\textwidth]{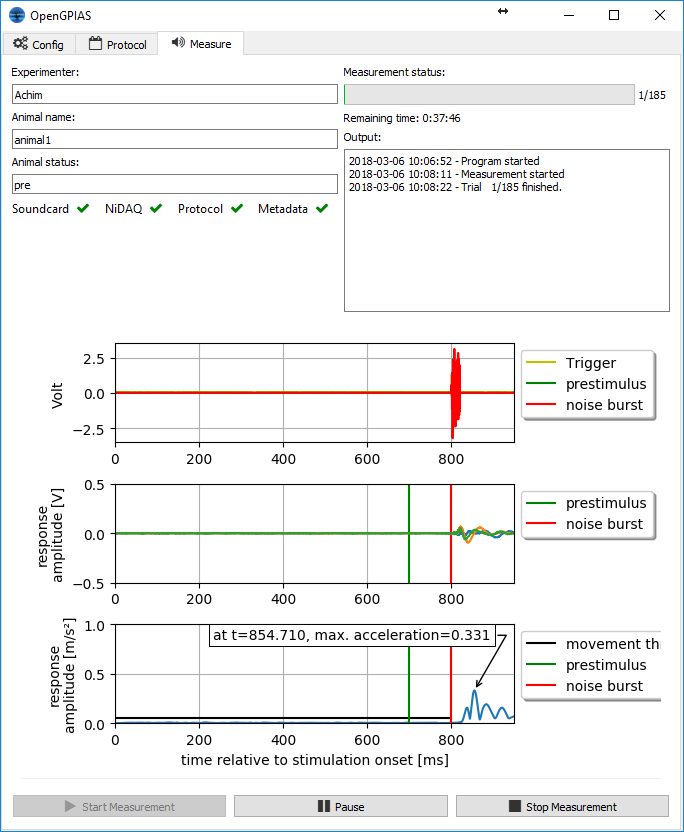}
\caption{\textbf{Measurement Interface}\newline
The interface of the measurement program. The different tabs provide access to the different modules. The measurement module provides inputs for the meta data, a status of the components, a log of the measurement, and a plot of the data from the last measurement.}
\label{fig:PLGUI}
\end{figure}

\subsection*{Measurement}
The main part of the software is the measurement module. This module checks whether both the soundcard and data acquisition card are configured properly and are ready to play or record, respectively. Furthermore, it checks the selected protocol file to be valid.

Before starting the measurement, the user has to supply some meta data, the experimenter name, the animal name, and optionally a treatment, to allow for consistent storage of the data.

When everything is ready and the animal has been placed in the restrainer, the measurement can be started. Each trial is presented to the animal using three channels of the sound card (trigger pulse, pre-stimulus, and startle-stimulus) and recorded using the data acquisition card (three input channels from the acceleration sensor and three input channels directly from the soundcard). Data is saved after each trial and raw-data is plotted to provide direct feedback to the experimenter.

The data is stored in a folder structure according to the meta data to allow for an organized storage. Data is saved in the form of numpy files (.npy) for space efficient storage, but can be exported to .txt or excel files for later evaluations.

\begin{figure*}[htb]
\centering
\includegraphics{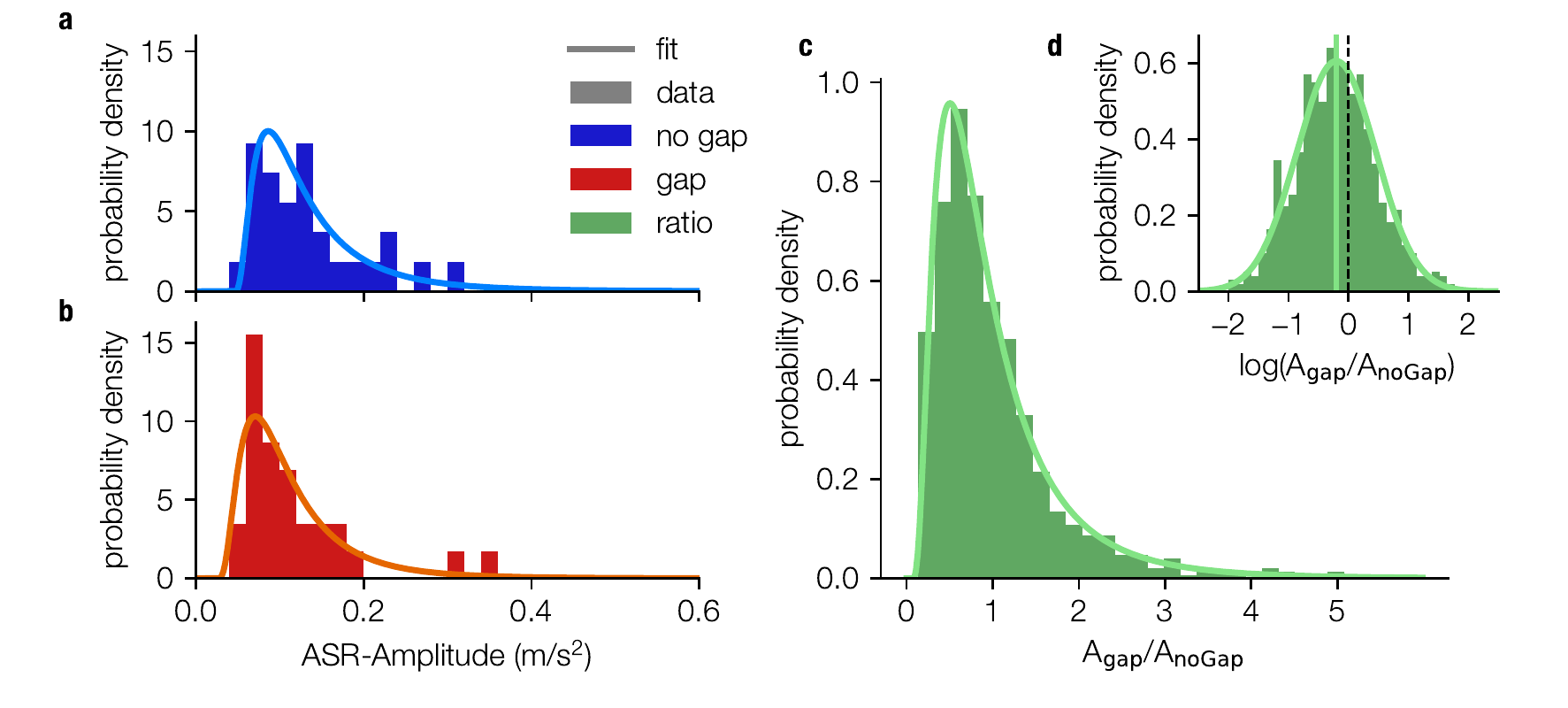}
\caption{\textbf{Distribution of ASR amplitudes.}\newline Distributions of the ASR amplitudes measured as a response to GPIAS paradigm stimuli, with a center frequency of 2\,kHz and a spectral width of $\pm 1/2$ octaves. \textbf{a}, ASR amplitudes without a gap of silence (blue). \textbf{b}, ASR amplitudes with gap stimuli (red). Both data sets are fitted with a lognormal distribution (solid line) using the maximum likelihood estimator. Histogram of the full combinatorial of the ratios of gap ($A_\mathrm{gap}$) and no gap ($A_\mathrm{nogap}$) ASR-amplitudes. \textbf{c}, Ratio histogram (green bars) can be described well with a lognormal distribution (solid line) as we have shown in a previous study \cite{Schilling2017}. \textbf{d}, The logarithmized ratios (green bars) are therefore Gaussian distributed (solid line). The mean of the distribution is negative, indicating that on average the pre-pulse inhibited the response.}
\label{fig:Amps}
\end{figure*}

\section*{Results}
In the following section, we show that exemplary recorded data using the novel setup is consistent with measurements from already established setups. 

Fig.\ \ref{fig:Amps} shows the ASR amplitudes for one exemplary animal stimulated according to the GPIAS paradigm (30 trials, center frequency 2,000\,Hz $\pm$ 1/2 octave). 

The data is in good agreement with results from a previous study conducted with another setup \cite{Schilling2017,Tziridis}, showing that ASR amplitudes are not normally distributed, but the startle responses can be well described by a lognormal distribution (Fig.~\ref{fig:Amps}). Also, the ratios of gap and no gap amplitudes are lognormally distributed (Fig.~\ref{fig:Amps}c), i.e. the logarithmic ratios are almost perfectly normally distributed (cf.~Fig.~\ref{fig:Amps}d).

The exemplary data clearly show that the here described measure for the ASR amplitudes based on the 3D acceleration vector (Equ.~\ref{eq:forcevector}) is a valid choice and leads to the same quantitative results as the 1D-peak-to-peak amplitudes used before. The average of the logarithm of the ratios is negative (Fig. \ref{fig:Amps}d) which means that the gap of silence leads to a pre-pulse inhibition, thus demonstrating the validity of the method.

\section*{Discussion}
In this study, we present an open source setup for the measurement of acoustic startle reflexes evoked by a loud noise burst with preceding pure-tone and/or gap pre-stimuli for the estimation of hearing thresholds and identification of possible tinnitus percepts in rodents. 

The pre-stimuli and startle stimuli (noise bursts) are presented using two separate loudspeakers, controlled by a commercial soundcard. The response amplitudes of the animals are captured by an 3D-acceleration sensor, whose output is digitized using a data acquisition card controlled via a Python program.

We have shown that the described setup can be used to present well defined pure-tone and band noise stimuli and to quantify the ASR amplitudes. Additionally, we were able to show that the pre-stimuli lead to a clear PPI (cf. Fig.~\ref{fig:Amps}c) and that the ASR amplitudes (cf. \cite{Csomor2008}) as well as the ratios of gap and no gap amplitudes  (cf. \cite{Schilling2017}) are lognormally distributed.

In contrast to existing, more elaborate and costly systems, our setup requires only low-cost hardware components and, with open-source software, allows for more flexible adoptions to different measurement protocols.

All our hardware components are listed in detail and the software is provided as open source (hardware and software documentation: \url{http://open-gpias.readthedocs.io}).

\FloatBarrier
\bibliographystyle{apalike}
\bibliography{Startle_Setup_References}

%\newpage
\end{document}